\documentclass[twocolumn,aps,prd,showpacs]{revtex4}
\usepackage{graphicx}
\usepackage{natbib}
\usepackage{amsbsy,amssymb,amsmath}
\usepackage{dcolumn}% Align table columns on decimal point
\usepackage[usenames]{color}
\usepackage{psfrag}
\usepackage[colorlinks,bookmarks]{hyperref}
\definecolor{linkblue}{rgb}{0,0,0.8}
\definecolor{linkgreen}{rgb}{0,0.5,0}
\hypersetup{pdfpagemode=UseNone, pdfstartview=FitH, linkcolor=linkblue, %
            citecolor=linkgreen, urlcolor=linkblue}

\newcommand{\beq}{\begin{equation}}
\newcommand{\eeq}{\end{equation}}

\def\cl{C_\ell}
\newcommand{\lp}{\left(}
\newcommand{\rp}{\right)}
\begin{document}

\title{Predicted Constraints on Cosmic String Tension from Planck and Future CMB Polarization Measurements}

\author{Simon Foreman}
%\email{sforeman@phas.ubc.ca}
\email{sfore@stanford.edu}
\affiliation{Department of Physics and Astronomy, %
University of British Columbia, %
Vancouver, BC, V6T 1Z1  Canada}

\author{Adam Moss} 
\email{adammoss@phas.ubc.ca}
\affiliation{Department of Physics and Astronomy, %
University of British Columbia, %
Vancouver, BC, V6T 1Z1  Canada}

\author{Douglas Scott} 
\email{dscott@phas.ubc.ca}
\affiliation{Department of Physics and Astronomy, %
University of British Columbia, %
Vancouver, BC, V6T 1Z1  Canada}

\date{\today}

\begin{abstract}
We perform a Fisher matrix calculation of the predicted uncertainties on estimates of the cosmic string tension $G\mu$ from upcoming observational data (namely, cosmic microwave background power spectra from the Planck satellite and an idealized future polarization experiment). We employ simulations that are more general than others commonly used in the literature, leaving the mean velocity of strings, correlation length of the string network, and ``wiggliness" (which parametrizes smaller-scale structure along the strings) as free parameters that can be observationally measured. In a new code, StringFast, we implement a method for efficient computation of the $\cl$ spectra induced by a network of strings, which is fast enough to be used in Markov Chain Monte Carlo analyses of future data. Performing a calculation with the string parameters left free results in projected constraints on $G\mu$ that are larger than those obtained by fixing their values {\it a~priori}, typically by a factor of $\sim$2--7. We also find that if $G\mu$ is equal to the current observational maximum, Planck will be able to make a confident detection of strings. However, if $G\mu$ is two orders of magnitude smaller, even a perfect, lensing-free measurement of polarization power spectra will not be able to detect a nonzero string tension at better than $2\sigma$ confidence.
\end{abstract}


\maketitle

%--------------------------------------------------------------------------------------
\section{Introduction}

In the last few decades, we have seen the arrival of an era of ``precision cosmology," as observational data have become precise enough to place serious constraints on our theories for the large-scale behaviour of the observable Universe. These data have led to the formulation of the standard model of cosmology (SMC; see, e.g., Ref.~\cite{scott-smc}), in which the Universe began in a hot, dense state and has been expanding and cooling for billions of years. However, the SMC is far from complete: there are mysterious substances, dark matter and dark energy, whose presence we can infer from data but have yet to fully understand, and there are also proposals for amendments to the model that incorporate as yet unobserved phenomena.

One such proposal is the existence of topological defects, especially cosmic strings (e.g.\ \cite{stringbook,kibble05,copeland-seekingstrings}). At one time, strings were thought to play an important role in generating structure in the Universe by seeding density perturbations at early times, but observations of the cosmic microwave background radiation (CMB) by the COBE satellite and later experiments contradicted this picture \cite{cobe-models,white-cobe,levon-defects}, instead favouring structure formation through quantum fluctuations that were amplified by a period of rapid inflation.

Nevertheless, cosmic strings have remained a subject of active study, since they are a generic prediction of superstring-inspired ``brane inflation" theories~\cite{sarangi-brane} and appear as a general feature of inflationary scenarios inspired by grand-unification--scale physics~\cite{strings-from-susy}. There are several proposed observational signatures of strings, including wedge-shaped wakes in 21cm redshift surveys~\cite{hernandez-21cmwake,brandenberger-21cmwake}, spatial correlations between anisotropies in 21cm radiation and the CMB~\cite{berndsen-21cmcorr}, gravitational waves from the decay of string loops \cite{vach-gwloop,depies-gwloop}, and gravitational lensing effects~\cite{vilenkin-strlens,sazhin-strlens,morganson-strlens,khlopov-lens}. Strings would also leave imprints on the CMB, such as step discontinuities in the temperature map~\cite{kaiser-stebbins} and B-modes of polarization \cite{levon-bmode,seljak-slosar-bmodes} (although the detection of B-modes may not be a ``smoking gun" for strings, or for any other theory for that matter---see Ref.~\cite{brandenberger-bmodes} for a recent discussion).

Because measurements of temperature anisotropies at small angular scales become dominated by the effects of hot gas in galaxy clusters and clustered infrared-emitting galaxies~\cite{birkinshaw-sz}, more precise measurements of polarization power spectra should be able to push cosmic string constraints to well below what is possible with temperature information alone. Indeed, several ongoing and upcoming projects (Planck~\cite{tauber-planck}, COrE~\cite{core}, ACTPol~\cite{actpol}, SPTPol~\cite{sptpol}, BICEP2/Keck~\cite{bicep2,keck}, POLARBEAR~\cite{polarbear}, and SPIDER~\cite{spider}, among others) will soon measure CMB polarization to unprecedented accuracy, and so in the near future we will be able to obtain strict limits on the cosmic string content of the Universe.

Consequently, this work proposes to use simulations of CMB polarization (and temperature, to a lesser extent) induced by cosmic strings to forecast the extent to which strings can be detected by upcoming measurements---in particular, by Planck or an idealized future measurement of polarization. Research that utilizes simulations of cosmic strings is ongoing \cite{cmbpol-strings,tensor-strings,battye-moss,bevis-ah,bevis2010,av-coupling}, and our contribution will be twofold.

First, in the spirit of ``precision cosmology," we will employ a more general model for strings than is commonly in use---namely, the one-scale model of, e.g., Ref.~\cite{wiggly}, but with the mean velocity, correlation length, and ``wiggliness" of the simulated network of strings left as adjustable parameters, in addition to the (dimensionless) string tension $G\mu$. Taking these properties as free, and in principle measurable from data, instead of fixed {\it a priori} leads to a weakening of projected constraints on $G\mu$, which we calculate using a Fisher matrix approach. We also develop a new code for quickly calculating the CMB spectra induced by a network of strings, which can be applied to the Markov Chain Monte Carlo (MCMC)~\cite{cosmomc} estimation of the one-scale string model parameters, and which will be made publicly available for use in other studies.

Second, we will perform forecasts both for a realistic observational scenario (CMB measurement by the Planck satellite) and a ``best-case scenario" situation in terms of detecting strings (noiseless measurement of polarization across a wide range of angular scales), incorporating the effects of gravitational lensing on the measurement. The latter will tell us how well we can ever do in constraining the string tension using CMB data, and also give an idea as to the range of string tensions we might be able to reliably discriminate from a Universe without strings ($G\mu=0$).

%--------------------------------------------------------------------------------------
\section{Modelling Cosmic Strings}
\label{sec:cmbstr}

The direct simulation of a network of cosmic strings without any simplifying assumptions is essentially impossible, as any such simulation would have to encompass an enormous range of physical scales to track the behaviour of strings from the early Universe until the present day. The standard approach uses the fact that, under some basic assumptions about string formation and decay, a network of strings evolves toward a scaling solution in which average properties of the network (such as the correlation length) scale in tandem with the expansion of the Universe. It is then possible to numerically solve either the Nambu or Abelian-Higgs (AH) equations of motion for a string, compute an unequal-time correlator of the stress-energy tensor of a simulated string network~\cite{pen-uetc}, and use this along with standard perturbative techniques to compute the resulting CMB anisotropies. However, Nambu and AH simulations generate networks with different correlation lengths and mean velocities (see Ref.~\cite{battye-moss} for details).

Another strategy is to represent a string network as a random collection of unconnected, straight segments, obtain the stress-energy tensor from these segments, and average the resulting CMB spectrum over many realizations of the network. This model has been implemented in the public code CMBACT~\cite{wiggly,cmbactsite}, which also utilizes the equation of state for ``wiggly" strings~\cite{carter-wiggly,vilenkin-wiggly}, incorporating the  uncertainty about small-scale structure along the strings via a ``wiggliness" parameter,~$\alpha$. The code calculates temperature anisotropies, $\cl^{\rm TT}$, as well as E- and B-modes of polarization, $\cl^{\rm EE}$ and $\cl^{\rm BB}$, and the cross power spectrum, $\cl^{\rm TE}$, including the effects from scalar, vector, and tensor components of the source function. We use CMBACT as the basis for the numerical calculations in this work. Note that CMBACT does not include a detailed treatment of production of small loops associated with string crossings, and also neglects the subsequent decay of these loops into gravitational radiation.

\begin{figure}
\centering \mbox{\resizebox{0.47\textwidth}{!}{\includegraphics[angle=0,trim=10 10 0 0]{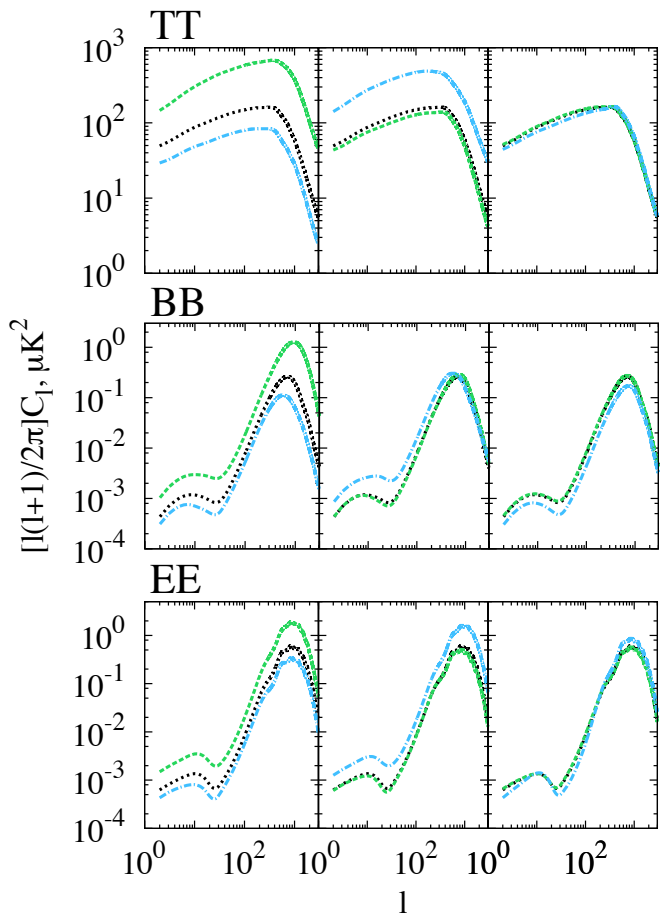}}}
\caption{Comparison of CMB power spectra (sum of scalar, vector, and tensor modes) when a single string parameter is varied while the others are kept constant. In each panel the black (dotted) line corresponds to a model with $G\mu=6.4\times 10^{-7}$, $\alpha=1.05$, $v=0.40$, and $\xi=0.35$. The leftmost frames show $\xi=0.15$ (green, dashed) and $\xi=0.5$ (blue, dot-dashed); the centre frames show $v=0.3$ (green, dashed) and $v=0.8$ (blue, dot-dashed); and the rightmost frames show $\alpha=1.0$ (green, dashed) and $\alpha=1.5$ (blue, dot-dashed). \label{fig:extremespecs}}
\end{figure}

The parameters $\xi$ (comoving correlation length of the network), $v$ (rms velocity of strings), and $\alpha$ take different values in the matter- and radiation-dominated eras, with their specific time evolution determined by CMBACT primarily from the effects of string expansion and loop production \cite{kibble1985,bennett1986,martins1996}. However, it is possible to explore a wider range of string properties if these three parameters are approximated as constant in time, so that they become free parameters of the model \cite{battye-moss,levon-bmode}, determined only by their initial conditions. Along with the string tension $G\mu$ (written as $G\mu/c^2$ with dimensions included), we then have a model for cosmic strings with four free parameters, that opens the door to such activities as MCMC estimation of these parameters or forecasts for how accurately they can be measured by future observations (this work is concerned with the latter endeavour, focusing on $G\mu$ in particular). We illustrate the effects of varying these parameters on CMB spectra in Fig.~\ref{fig:extremespecs}.

It has already been shown that these parameters can be chosen to give spectra that match the output of Nambu and AH simulations to a reasonable level of accuracy \cite{battye-moss}. While it is true that future string calculations may be able to include a wider range of detailed physical effects, we believe that our four-parameter approach at least has the virtue of being more realistic that the one-parameter assumption that is commonly adopted.

%--------------------------------------------------------------------------------------
\section{stringfast: A Code for Fast String Calculations}

For a single set of string parameters, using CMBACT to compute the resulting CMB spectra can take anywhere from several hours to several days on a single modern CPU. The required time depends on the number of realizations of the string network, which is directly related to the accuracy of the $\cl$s generated as output, since the final $\cl$s are averages over all realizations. Clearly, these computations must be optimized significantly before applications that require large numbers of $\cl$ calculations, such as MCMC studies, can become feasible.

\subsection{Strategy}

It has been proposed that ``morphing" techniques borrowed from computer graphics could be used for efficient calculations of CMB spectra \cite{sigurdson-morph}. This morphing approach defines certain Òcontrol pointsÓ based on special features of the spectra (such as acoustic peaks and troughs), interpolates between pre-calculated curves to determine where these points would be for a different set of input (e.g.\ cosmological) parameters, and then smoothly transforms the two adjacent curves to a new curve that preserves the key features of the original curves.

\begin{figure}
\centering \mbox{\resizebox{0.47\textwidth}{!}{\includegraphics[angle=0,trim=10 10 0 0]{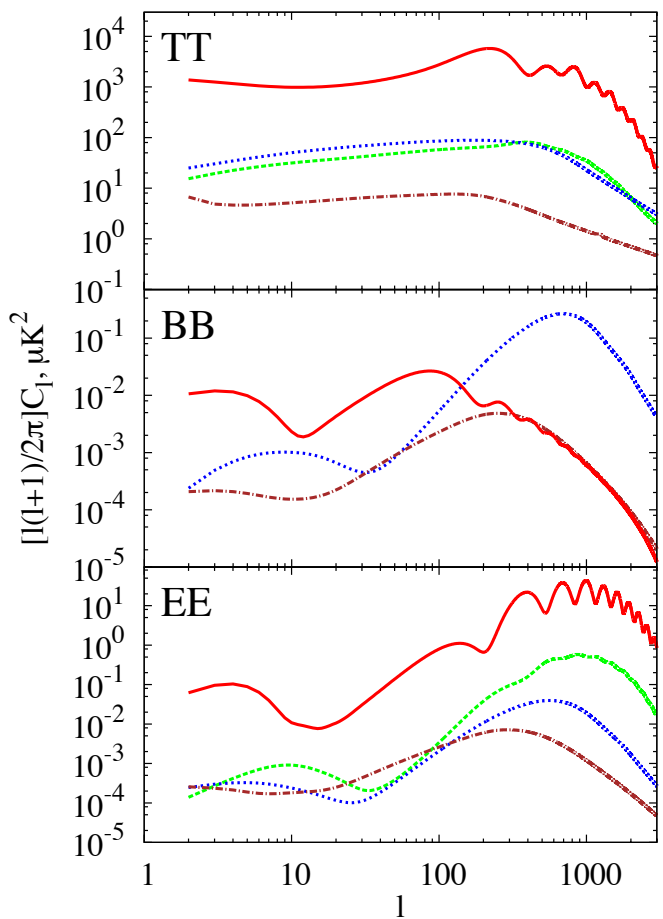}}}
\caption{Scalar (green, dashed), vector (blue, dotted), and tensor (brown, dot-dashed) modes of CMB power spectra generated by strings with $G\mu=6.4\times 10^{-7}$, $\alpha=1.05$, $v=0.40$, and $\xi=0.35$. The sum of inflationary scalar, vector, and tensor modes generated by CAMB, using WMAP7 parameters and $r=0.36$, are also shown (red, solid). \label{fig:examplespecs}}
\end{figure}

The spectra generated by cosmic strings are dominated by only a small number of key features, such as a low-$\ell$ bump from reionization and a larger peak between $\ell\approx 500$ and $1000$ from recombination (see Fig.~\ref{fig:examplespecs}). Therefore, it is unnecessary to implement a full morphing scheme---rather, the spectra can be described by a small number of fitting functions, or, in cases where there is nontrivial variation of a spectrum with respect to string network parameters, by cubic splines. There is then a straightforward strategy for calculating the spectra for arbitrary values of the parameters (similar to what is done in some inflationary CMB codes, such as DASh~\cite{dash} and Pico~\cite{pico}):
\begin{enumerate}
\item Pre-calculate the spectra on a grid in parameter space, with limits based on reasonable priors for the parameters and spacing chosen to give acceptable accuracy in the approximate curves generated by the method.
\item Fit each pre-calculated spectrum with fitting functions or cubic splines.
\item Create multi-dimensional splines, as functions of the model parameters, for each fit parameter (or, if a curve is fit by splines, for each $\cl$ value used in that fit).
\item For the desired set of model parameters, use these splines to calculate the corresponding fit parameters (or spline control points), and use these to output the resulting spectra.
\end{enumerate}

It is possible to encapsulate this process into a distributable Fortran module, which we dub StringFast. From an end-user perspective, the first two steps will have been completed beforehand, and files of fit parameters and/or spline control points will be distributed along with the source code. At runtime, one first calls subroutines which read in these files and assemble the multi-dimensional splines into memory. One can then use the main routines of the module, which perform step 4, to calculate a specific $\cl$ value as a function of spectrum type (TT, EE, or BB, as well as scalar, vector, or tensor), $\ell$, string model parameters, and optical depth $\tau$ of the background cosmology (see below). The resulting time for StringFast to compute a complete ($2<\ell<3000$) $\cl$ curve is less than a second on a modern CPU.

In addition to the four parameters of the one-scale string model, we must account for how the string spectra change with variation of the optical depth $\tau$, since this value has an important effect on the height of the reionization bump, and also scales the height of the main peak by a factor of $\sim\! e^{-2\tau}$ (as roughly $e^{-\tau}$ of the photons are scattered on their way from the last-scattering surface). For the other cosmological parameters, however, we assume that variations of their values consistent with current observational constraints only have higher-order effects (that can be safely neglected) on the $\cl$s generated by strings. The code evaluates the string spectra using the best-fit parameter values from the WMAP seven-year data analysis \cite{larson-wmap} (although this could be changed by running a new grid of spectra).

\subsection{Accuracy}

For the spectra pre-computed on the grid in parameter space, we run CMBACT with 12,000 network realizations. To determine the precision of these spectra, we choose several sets of sample parameters, generate eight sets of 12,000-realization curves, and calculate the ideal signal-to-noise (S/N) of the maximum {\it difference} signal. Specifically, we calculate the maximum value of $\lp \cl^A - \cl^B \rp / \Delta\cl^{AB}$ over $2<\ell<3000$ and over all pairs $(A,B)$ of our eight test curves, where $\Delta\cl^{AB}=\sqrt{2/(2\ell+1)} \lp \cl^A+\cl^B \rp$ (the ``noise" being merely the cosmic variance of the difference signal). This measures the uncertainty of a simulated curve at any point along the curve, as compared to the noise inherent in a measurement of the $\cl$s.

The maximum S/N we find is 1.40, and in fact most types of spectra have maximum S/N $<1$. Thus, the variation in the pre-computed spectra does not rise significantly above the fundamental uncertainty of power spectrum observations. Moreover, since the underlying model we employ is itself only an approximation of the more intricate physics of cosmic strings (ignoring, for example, the details of the production of loops), we are confident that this is an adequate level of accuracy for the forecasts we wish to undertake, and also for more general studies of the average properties of a network of strings.

To assess the accuracy of the approximate (off-grid) spectra generated by StringFast, we compute a $\chi^2$ statistic comparing the approximate curves to ``exact" (96,000-realization) curves generated by CMBACT with the same parameter values:
\beq
\label{eq:ch2}
\chi^2 = \sum_{\ell=2}^{3000} \frac{\lp \cl^{\rm appr} - \cl^{\rm CMBACT}\rp^2}{\sigma_{\cl}^2},
\eeq
where the $\sigma_{\cl}$ are computed from eight sets of test curves and their average. For a selection of test parameters, approximate curves are generated using a pre-computed grid described in Tables \ref{tab:pargrid1}, with resulting reduced $\chi^2$s given in Table \ref{tab:redch1}. Since the $\cl$s depend on $G\mu$ through the simple analytical relationship $\cl\propto (G\mu)^2$, this parameter is not included in the grid. The range of parameters in Table \ref{tab:pargrid1} is sufficient for Fisher matrix calculations using the fiducial model of Section~\ref{sec:fidmod}; future versions of the code will cover a wider region of parameter space.

\begin{table}[t]
\begin{center}
\begin{tabular}{|l||l|l|l|}

	\hline \textbf{parameter} & \textbf{min.\ value} & \textbf{max.\ value} & \textbf{step size} \\ \hline\hline
      $\tau$ & 0.085 & 0.095 & 0.005 \\
      $\alpha$ & 1.0 & 1.5 & 0.1 \\
      $v$ & 0.30 & 0.80 & 0.05 \\
      $\xi$ & 0.10 & 0.50 & 0.05 \\
      \hline 
\end{tabular}
\caption{\label{tab:pargrid1} Attributes of grid used for accuracy tests in Table~\ref{tab:redch1}.}
\end{center}
\end{table}

\begin{table}
\begin{center}
\begin{tabular}{|l|l|l|l||r|r|r|r|r|r|r|r|}

	\hline \multicolumn{4}{|c||}{{ }} & \multicolumn{8}{|c|}{\textbf{Reduced $\chi^2$, comparing }} \\
	\multicolumn{4}{|c||}{{ }} & \multicolumn{8}{|c|}{\textbf{approximate curve to}} \\
	\multicolumn{4}{|c||}{{ }} & \multicolumn{8}{|c|}{\textbf{``exact" curve}} \\
	\hline \multicolumn{4}{|c||}{\textbf{input parameters}} & \multicolumn{3}{|c|}{\textbf{TT}} & \multicolumn{2}{|c|}{\textbf{BB}} & \multicolumn{3}{|c|}{\textbf{EE}} \\
	\hline
	 $\boldsymbol{\tau}$ &  $\boldsymbol{\alpha}$ & $\boldsymbol{v}$ & $\boldsymbol{\xi}$ & {\bf s} & {\bf v} & {\bf t} & {\bf v} & {\bf t} & {\bf s} & {\bf v} & {\bf t} \\ \hline      \hline
	 
	 0.087 & 1.26 & 0.58 & 0.19 &	5.5 & 64.1 & 11.0 & 10.7 & 8.6 & 2.0 & 23.6 & 6.0 \\
	 0.088 & 1.05 & 0.40 & 0.35 &	3.4 & 1.3 & 2.9 & 10.1 & 21.9 & 5.2 & 8.2 & 17.5 \\
	 0.090 & 1.20 & 0.77 & 0.27 &	6.9 & 1.1 & 2.0 & 3.7 & 14.1 & 5.3 & 3.6 & 8.9 \\
	 0.090 & 1.30 & 0.40 & 0.23 &	9.0 & 26.9 & 0.3 & 3.3 & 3.4 & 1.1 & 5.6 & 3.5 \\
	 0.091 & 1.08 & 0.78 & 0.44 &	9.7 & 8.0 & 2.2 & 8.8 & 5.7 & 6.6 & 7.7 & 3.7 \\
	 0.091 & 1.15 & 0.33 & 0.42 &	3.5 & 1.1 & 2.4 & 2.0 & 3.6 & 13.5 & 1.9 & 3.6 \\
	 0.092 & 1.20 & 0.47 & 0.28 &	5.8 & 4.3 & 4.9 & 6.5 & 2.0 & 5.8 & 9.5 & 3.5 \\
	 0.094 & 1.43 & 0.51 & 0.33 & 	1.3 & 1.6 & 1.3 & 0.9 & 1.9 & 2.2 & 0.9 & 0.9 \\
	 
	 \hline
	 
\end{tabular}
\end{center}
\caption{\label{tab:redch1} Reduced $\chi^2$ (i.e.\ $\chi^2/3000$) values for approximate spectra generated from the specified input parameters, using the grid described in Table~\ref{tab:pargrid1}. We report results for scalar, vector, and tensor modes, denoted by ``s," ``v," and ``t" respectively.}
\end{table}

It is important to realize that the $\chi^2$ statistic above is not necessarily the most useful or informative way of assessing the accuracy of the approximate curves. One reason is that for some spectra, the major contribution to the $\chi^2$ values comes from a certain range in $\ell$. For example, for BB tensors in the second line of Table \ref{tab:redch1}, it is only the $\cl$s for $\ell >1100$ that display significant deviations from the exact curve. However, ignoring the range $2000<\ell <3000$ in our Fisher matrix calculations only affects our forecasts at the $\sim\!\! 10\%$ level, so in this case the sum in Eq.~(\ref{eq:ch2}) runs over angular scales that are not significant for this application.

Also, the accuracy of approximate curves is highly dependent on the spacing and extent of the pre-computed grid of models. Thus, in regions of the grid where the approximate curves are less accurate (our tests revealed $\xi<0.15$ to be one such region, if the grid spacing in $\xi$ is 0.05 as in Table \ref{tab:pargrid1}), the results can be improved by refining the grid and introducing more pre-computed spectra in these regions. Such refinements of the grid entail a trade-off between accuracy of the output and memory required at runtime, but the latter is not expected to present significant obstacles to obtaining usable spectra from the code. Future versions of the code, making use of larger and finer grids, will be able to produce more accurate curves where necessary.

Another reason to be cautious in interpreting the tables above is that they compare the accuracy of the approximate curves to the variation in the ``exact" curves from CMBACT (run at a high number of realizations), but it is likely untrue that $\chi^2_{\rm red}<1$ is necessary for effective use of the spectra, given the shortcomings of the model for strings on which CMBACT is based. In addition, even the ``exact" curves will have some deviation, so a comparison to these curves is not completely equivalent to a comparison to the idealized limiting curves that are ostensibly approached as the number of realizations is taken to be arbitrarily large.

Thus, it is recommended that the accuracy of StringFast's output should be assessed on an application-by-application basis, with due consideration of the precision that might be desired in various ranges of $\ell$. In using StringFast to compute numerical derivatives of $\cl$s with respect to the parameters of the model, we assert that the accuracy of the approximate curves near our chosen fiducial model (the second line of Table \ref{tab:redch1}) is sufficient, since small and non-systematic errors in the curves will in general be washed out by the sum over $\ell$ in the computation of the Fisher matrix [Eq.~(\ref{eq:fish-wcov})].

%--------------------------------------------------------------------------------------
\section{Methods}

\subsection{Fisher Matrices}
\label{sec:fisher}

We obtain projected uncertainties on determinations of the string model parameters from the Cram\'{e}r-Rao inequality~\cite{kendallstewart}: if $\hat{\lambda}_i$ is an unbiased estimator for a parameter $\lambda_i$, then
\beq
{\rm Var}\!\lp\hat{\lambda}_i\rp \geq \lp \mathbf{F}^{-1} \rp_{ii},
\eeq
where the Fisher matrix $\mathbf{F}$ for CMB observations is given by
\beq
\label{eq:fish-wcov}
\mathbf{F}_{ij} = \sum_\ell \sum_{X,Y} \frac{\partial\cl^X}{\partial\lambda_i} {\rm \mathbf{Cov}}^{-1}(\cl^X\cl^Y)  \frac{\partial\cl^Y}{\partial\lambda_j},
\eeq
with $X$, $Y$ standing for TT, EE, BB, or TE, and the covariance matrix $\mathbf{Cov}$ incorporating signal and noise variance on measurements of the $\cl$s (see, e.g., Ref.~\cite{eisenstein-fisher}). Thus, we take $\sqrt{(\mathbf{F}^{-1})_{ii}}$ as the expected uncertainty on a measurement of parameter $\lambda_i$.

We calculate the derivatives in Eq.~(\ref{eq:fish-wcov}) numerically via finite-differencing, using StringFast (with the grid described in Table~\ref{tab:pargrid1}) to calculate the string $\cl$s and the public Boltzmann code CAMB~\cite{camb} for the inflationary $\cl$s. Gravitational lensing of inflationary E-modes into B-modes is accomplished by CAMB; in the forecasts where lensing is not removed from the final signal, the lensed B-modes are taken as contributing to the experimental noise in the covariance matrix. Since the E-modes generated by strings are $\sim$2 orders of magnitude weaker than those generated by inflation (see the bottom panel of Fig.~\ref{fig:examplespecs}), lensing of these modes is ignored in our calculation.

We investigate the prospects for constraining the properties of strings using the Planck satellite, and also using some idealized, cosmic variance-limited measurement of CMB polarization, which we refer to as ``FuturePol" for brevity. Since the main constraints on strings will come from B-modes of polarization, this allows us to obtain the most optimistic estimate of how precisely we can hope to detect strings via their signatures in the CMB.

For Planck, we use characteristics of the lowest six frequency channels as presented in~\cite{tauber-planck}, combining the noise from all six channels via
\beq
\label{eq:nl-ch}
N_\ell^Z = \left[ \sum_c \frac{1}{\theta_c^2 \lp\Delta_c^Z\rp^2} \, {\rm exp} \! \left( -\frac{\ell^2\theta_c^2}{8\, {\rm ln} \, 2} \right)  \right]^{-1},
\eeq
where $c$ runs over the channels, $Z$ denotes either temperature or polarization, $\theta_c$ is the beam full-width-half-max, and $\Delta_c^Z$ is the experimental noise per pixel (see, e.g., Ref.~\cite{efstathiou-nl}). We use all four types of $\cl$s (TT, EE, BB, and TE) in the Fisher matrix.

For FuturePol, we assume the ideal case of no experimental noise. We utilize only EE and BB, motivated by the idea that future polarization measurements will surpass the sensitivity threshold of temperature anisotropies, because of the lack of significant small-scale foregrounds for polarization signals. However, in order to account for the fact that an experiment like FuturePol would already have the Planck data at its disposal when the observations are analyzed, we use the constraints from the Planck Fisher calculation as Gaussian priors on each parameter. This is accomplished by adding $1/\sigma^2$ to the corresponding off-diagonal element of the FuturePol Fisher matrix, where $\sigma^2$ is the forecast experimental variance on each parameter obtained from $\mathbf{F}^{-1}$ for Planck \cite{lsst-book}.

\subsection{Fiducial Model}
\label{sec:fidmod}

For the fiducial model we use in our forecasts (which then determines the point in parameter space where Eq.~(\ref{eq:fish-wcov}) is evaluated), we choose model parameters that were found in Ref.~\cite{battye-moss} to give spectra that agree well with the results of field-theoretic simulations for an AH string model \cite{bevis-ah}: $\alpha=1.05$, $v=0.40$, and $\xi=0.35$.

For the string tension $G\mu$, we forecast for three values: the current observational maximum for AH strings, calculated from current CMB data and galaxy surveys \cite{battye-moss} to be $6.4\times 10^{-7}$; and reductions of the tension by one and two orders of magnitude, $6.4\times 10^{-8}$ and $6.4\times 10^{-9}$. The first case allows us to investigate the ``best-case scenario" for the detection of strings, in which the string contribution to the CMB is as high as possible, while the other cases illuminate our prospects for detecting less-energetic strings. The background cosmology is described using six parameters ($\Omega_{\rm b} h^2$, $\Omega_{\rm c} h^2$, $H_0$, $\Delta_\mathcal{R}^2$, $n_{\rm s}$, and $\tau$) with values taken from the WMAP seven-year data analysis \cite{larson-wmap}.

\begin{table}[t]
\begin{center}
\begin{tabular}{|D{.}{.}{5}||D{.}{.}{3.3}|D{.}{.}{1.3}|D{.}{.}{3.3}|D{.}{.}{1.3}|}
\hline
	& \multicolumn{4}{c|}{{$\boldsymbol{\sigma}_{\mathbf{G}\boldsymbol{\mu}}\boldsymbol{\times} \mathbf{10^{7}}$}} \\
	\hline
	& \multicolumn{2}{c|}{{\textbf{$\mathbf{G}\boldsymbol{\mu}$ only}}}
	& \multicolumn{2}{|c|}{{\textbf{$\mathbf{G}\boldsymbol{\mu}$ + 3 others}}} \\
	\hline
	\multicolumn{1}{|c||}{{\textbf{$\mathbf{G}\boldsymbol{\mu}\boldsymbol{\times} \mathbf{10^{7}}$}}} & 
	\multicolumn{1}{|c|}{{Planck}} & \multicolumn{1}{|c|}{{FuturePol}} &
	\multicolumn{1}{|c|}{{Planck}} & \multicolumn{1}{|c|}{{FuturePol}} \\
	\hline
      6.40 & 0.15 & 0.018 & 0.64 & 0.073   \\
      0.64 & 1.25 & 0.053 & 5.98 & 0.35   \\
      0.064 & 12.5 & 0.52 & 59.8 & 3.41 \\
      \hline
\end{tabular}
\caption[Forecast ($1\sigma$) uncertainties on estimations of $G\mu$ from data.]{\label{tab:gmuconlens} Forecast ($1\sigma$) uncertainties on estimations of $G\mu$ from data, with no removal of lensing from the B-mode signal. The titles ``$G\mu$ only" and ``$G\mu$ + 3 others" refer to the string model parameters that are left free in the Fisher matrix. For the latter (more realistic) case, a non-zero value of $G\mu$ cannot be detected at better than $2\sigma$ confidence if $G\mu\lesssim 6.4\times 10^{-8}$.}
\end{center}
\end{table}

\begin{table}
\begin{center}
\begin{tabular}{|D{.}{.}{5}||D{.}{.}{3.3}|D{.}{.}{1.3}|D{.}{.}{3.3}|D{.}{.}{1.3}|}
\hline
	& \multicolumn{4}{c|}{{$\boldsymbol{\sigma}_{\mathbf{G}\boldsymbol{\mu}}\boldsymbol{\times} \mathbf{10^{7}}$}} \\
	\hline
	& \multicolumn{2}{c|}{{\textbf{$\mathbf{G}\boldsymbol{\mu}$ only}}}
	& \multicolumn{2}{|c|}{{\textbf{$\mathbf{G}\boldsymbol{\mu}$ + 3 others}}} \\
	\hline
	\multicolumn{1}{|c||}{{\textbf{$\mathbf{G}\boldsymbol{\mu}\boldsymbol{\times} \mathbf{10^{7}}$}}} & 
	\multicolumn{1}{|c|}{{Planck}} & \multicolumn{1}{|c|}{{FuturePol}} &
	\multicolumn{1}{|c|}{{Planck}} & \multicolumn{1}{|c|}{{FuturePol}} \\
	\hline
      6.40 & 0.15 & 0.0038 & 0.62 & 0.028   \\
      0.64 & 1.14 & 0.0021 & 5.74 & 0.0070   \\
      0.064 & 11.4 & 0.018 & 57.4 & 0.039 \\
      \hline
\end{tabular}
\caption{\label{tab:gmuconnolens} As Table \ref{tab:gmuconlens}, but with 100\% efficiency in removing the lensing B-mode signal. Even without contamination from lensing, in the ``$G\mu$ + 3 others" case $G\mu$ cannot be distinguished from zero at better than $2\sigma$ confidence if $G\mu\lesssim 6.4\times 10^{-9}$.}
\end{center}
\end{table}

Previous work on strings (e.g.\ \cite{cmbpol-strings}) has not taken $\alpha$, $v$, and $\xi$ as measurable quantities, instead making {\it a~priori} assumptions about the corresponding properties and therefore the shape of the resulting power spectrum. Thus, to compare the previous approach with the current work, we perform forecasts in two ways: first, letting $\alpha$, $v$, and $\xi$ be free, adjustable parameters (with corresponding elements in the Fisher matrix), and second, fixing them at their AH values and not including them in our Fisher matrix calculations.

%--------------------------------------------------------------------------------------
\section{Results}

\begin{figure}
\centering \mbox{\resizebox{0.49\textwidth}{!}{\includegraphics[angle=0,trim=5 10 0 0]{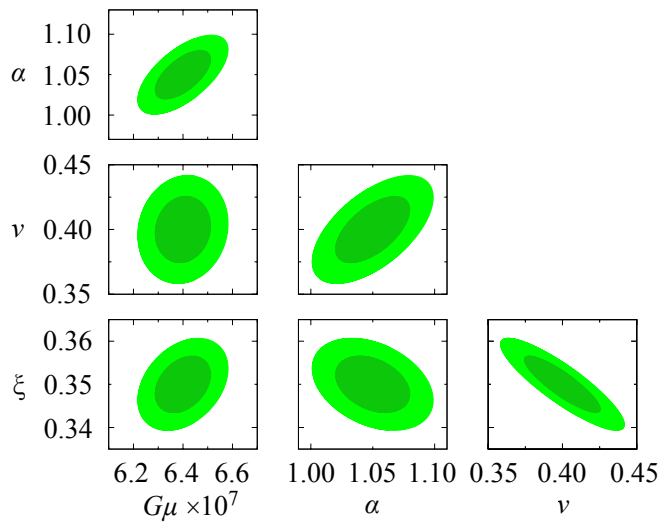}}}
\caption{Prediction ellipses for string parameters measured by FuturePol, with $G\mu=6.4\times 10^{-7}$ and no lens cleaning. Dark green (inner) regions denote forecasts at 68\% C.L., while light green (outer) regions correspond to 95\% C.L. While some degeneracies are present, they are not severe enough to impede accurate estimation of parameter values from future data. \label{fig:els}}
\end{figure}

The forecast $1\sigma$ uncertainties on estimations of $G\mu$ from future data are shown in Table \ref{tab:gmuconlens}, where the lensing B-mode signal is used as additional noise in the covariance matrix, and Table \ref{tab:gmuconnolens}, where we assume that the entire lensing signal can be subtracted. There are several notable features of these results. First, moving from the ``$G\mu$ only" model to the (more realistic) ``$G\mu$ + 3 others" model increases the uncertainty on estimates of $G\mu$ by a factor $\sim$2--7. Since the three other string parameters can be estimated from CMB observations, it is therefore important that they be allowed to vary in a thorough search for signatures of strings---fixing their values {\it a~priori} could lead to erroneously precise determinations of the string tension. Fig.~\ref{fig:els} shows anticipated correlations between measurements of the parameters; several parameters are weakly degenerate due to certain similar effects they have on CMB spectra (e.g.\ scaling the overall amplitude).

Second, if the ``$G\mu$ + 3 others" analysis is used, and if the string tension is equal to the AH observational maximum, then Planck should have sufficient precision in its CMB measurements to determine that the string tension is non-zero at high statistical significance. Thus, it will be able to make a definitive detection of the presence of strings in the Universe (provided they are described by the AH model we forecast for here). However, if the string tension is just one order of magnitude below the observational maximum, then the efficiency with which the effects of lensing can be removed from the B-mode signal impacts the constraints on $G\mu$. With 100\% removal, FuturePol can measure the tension very precisely, while with no removal, it cannot discriminate between the presence ($G\mu\neq 0$) and absence ($G\mu=0$) of strings at better than $\sim$$2\sigma$ confidence. For $G\mu$ two orders of magnitude below its upper bound, even FuturePol with 100\% lensing removal cannot do any better than a $\sim$$2\sigma$ detection of a non-zero $G\mu$ through its measurement of the $\cl$s.

We can compare this result to a recent similar study~\cite{cmbpol-strings}, which asserts that AH-generated strings can be detected by CMBPol \cite{cmbpol} at $3\sigma$ confidence if $G\mu\simeq 9\times 10^{-8}$. That work uses the cosmic variance of the weak lensing B-modes as additional Gaussian noise, while we use either the entire lensing signal as noise or assume that lensing can be completely removed in the final analysis. We agree that after assuming specific shapes for the $\cl$ curves, so that the only freedom left is to scale the overall amplitudes via $G\mu$, CMBPol or our FuturePol will be able to detect a nonzero string tension at high confidence, even without removal of any contamination from lensing (see the ``$G\mu$ only" section of Table \ref{tab:gmuconlens}).

However, we have shown that there are two factors that can strongly impact the precision with which we can measure $G\mu$: the efficiency of lensing removal, and the way the properties of strings are allowed to change the shape of the CMB power spectrum. We reiterate that it is preferable to relinquish the assumption of a specific shape for the spectrum (an effective fixing of $\alpha$, $v$, and $\xi$) and allow other properties of strings to be measured from observations, in which case strings with $G\mu\lesssim7\times10^{-9}$ cannot be reliably constrained even by a noiseless, lensing-free measurement of polarization power spectra. Thus, theories that predict a wide possible range of orders of magnitude for $G\mu$ will require observational probes other than the CMB $\cl$s to be verified or ruled out completely.

%--------------------------------------------------------------------------------------
\section{Conclusions}
\label{sec:conc}

Our primary goal in this paper has been to predict how accurately the tension of cosmic strings can be measured by upcoming measurements of CMB polarization (and temperature, to a lesser extent). To do this, we have developed a new scheme to quickly calculate the $\cl$s from a network of strings with certain specified properties (tension, mean velocity, correlation length, and ``wiggliness"), and implemented this scheme in a modular code that we call StringFast. Our calculations have the advantage of incorporating these properties as adjustable parameters of the model for strings, as opposed to fixing these properties {\it a priori} by assuming a specific shape for the power spectrum generated by strings, which is often done in the literature. Using a Fisher matrix approach, we have estimated how confidently we can constrain $G\mu$ with upcoming CMB data, and also how much this constraint is weakened by adding additional physically-motivated free parameters to the numerical description of strings that is employed.

Our results highlight the importance of prior assumptions in the search for cosmic strings in observational data. It is advisable to allow for more general string properties than those of any one field-theoretic model or numerical implementation, and thereby incorporate the uncertainty inherent in computational approaches to modelling strings---but, doing so also weakens our ability to detect strings via their effects on the CMB. Our forecasts also indicate that theories that generate strings with $G\mu\ll10^{-9}$ do not have strong prospects for verification via CMB power spectra, and thus we may need to look to gravitational wave signals or other signatures for empirical tests. Note, however, that our analysis only applies to field-theoretic strings and not to large-scale superstrings; see Ref.~\cite{av-coupling} for a similar recent study of superstring networks with multiple tensions and Y-junctions.

With the impending release of data from Planck, and with other upcoming probes of polarization, we will soon be able to place new, tighter constraints on the energy scale and properties of cosmic strings. While the search for strings is not the primary aim of these experiments, their data can and should be used for an attempted detection (or null detection) of strings, either of which would provide important insights into the physics of the early Universe.

%----------------- ACKNOWLEDGMENTS -----------------------

\section*{Acknowledgments}

This research was supported by the Natural Sciences and Engineering Research Council of Canada and the Canadian Space Agency. Calculations were performed on the Sunnyvale cluster (funded by the Canada Foundation for Innovation) at the Canadian Institute for Theoretical Astrophysics. We thank Levon Pogosian for useful discussions regarding CMBACT.

\bibliography{strings_bib}

\end{document}